\documentclass[10pt,conference]{IEEEtran}
\usepackage[utf8]{inputenc}

\IEEEoverridecommandlockouts

\usepackage{graphicx}
\usepackage[utf8]{inputenc}
\usepackage{latexsym}
\newcommand{\B}{$\Box$}

\newcommand{\hide}[1]{}
\newcommand{\code}[1]{\textsf{#1}}
\newcommand{\anon}[2]{#1}

\begin{document}

\title{Considerations and Pitfalls in\\ Controlled Experiments on Code Comprehension%
\thanks{\anon{Dror Feitelson holds the Berthold Badler chair in Computer Science. This research was supported by the ISRAEL SCIENCE FOUNDATION (grant no.\ 832/18).}{}}
}

\author{\anon{\IEEEauthorblockN{Dror G. Feitelson}
\IEEEauthorblockA{Department of Computer Science\\
The Hebrew University of Jerusalem, 91904 Jerusalem, Israel}}
{Authors Anonymized}}

\maketitle

\begin{abstract}
Understanding program code is a complicated endeavor.
As such, myriad different factors can influence the outcome.
Investigations of program comprehension, and in particular those using controlled experiments, have to take these factors into account.
In order to promote the development and use of sound experimental methodology, we discuss potential problems with regard to the experimental subjects, the code they work on, the tasks they are asked to perform, and the metrics for their performance.
\end{abstract}

\section{Introduction}

Code comprehension is a major element of software development.
According to Robert Martin, developers read 10 times more code than they write \cite{martin:clean}.
In one survey, 95\% of developers said understanding code was an important part of their job, and a large majority said they do it every day \cite{cherubini07}.
Most developers also agree that understanding code written by others is hard.
As researchers, we are interested in exactly what makes it hard, and what can be done about it.

Controlled experiments are at the heart of research on code comprehension \cite{weissman74,siegmund16}.
In such experiments experimental subjects are asked to perform a programming task on given code.
The task is crafted so that performing it successfully requires the code to be understood.
By measuring the effort and success of performing the task, one can therefore obtain some information on the difficulty of understanding the code.
Repeating the measurements on modified code or using various tools then sheds light on the effect of code features and tools on program comprehension.

While the general framework of code comprehension experiments is well known, there are many variations in the details.
This is a natural result of the combination of the many decisions that have to be taken:
\begin{itemize}
    \item One has to select the code on which the subjects will work.
    The code often reflects the nature of the study, e.g.\ using a certain style of identifiers if the effect of such styles on comprehension is the focus of the study.
    However, many other attributes of the code may also affect the comprehension process.
    It is therefore important to select code that does not introduce threats to the validity of the study.
    \item One has to select the tasks to be performed.
    The tasks are supposed to require understanding, but what does ``understanding'' mean?
    A particular risk is that subjects may be able to find shortcuts and perform the task without actually understanding the code, thereby undermining the whole experiment.
    \item One has to select the metrics by which performance will be measured.
    Different metrics may actually measure different things, and reflect different aspects of the difficulties in understanding code.
    \item One also has to select the subjects themselves.
    A much-cited threat is the use of students as experimental subjects.
    But when are students indeed a problem, and when can they be used safely?
    And is the student/professional dichotomy indeed the correct one to be concerned about?
\end{itemize}

Most work on the methodology of empirical software engineering focuses on experimental design, statistical tests, and reporting guidelines \cite{basili86,juristo:exp,wohlin:exp,shull:exp}.
But it is also important to get the domain-specific core features right \cite{brooksre80}.
For program comprehension our focus will be on general attributes of the code, the tasks, and the measurements.
We will not discuss the manipulations that are part of a specific research agenda.
Our goal is to review the choices that have been made in experimental studies, and the risks involved in them.
Hopefully this will encourage additional work on the methodological aspects of code comprehension research.
The current selection of what to discuss is based on personal experience.
It should be considered as a basis for discussion, not a comprehensive listing.

Some previous work in this domain includes the following.
In an early work Brooks identified four factors which affect comprehension: what the program does, the program text, the programmer's task, and individual differences \cite{brooksr83}.
These foreshadow some of our observations on the code, the task, and the experimental subjects.
Littman et al.\ defined understanding at a somewhat higher level of abstraction \cite{littman87}.
According to them, understanding a program comprises knowing the objects the program manipulates and the actions it performs, as well as its functional components and the causal interactions between them.
Note that this description relates to systems, and not to smaller elements of code.

Perhaps the closest to our work is Siegmund's review of confounding parameters in program comprehension \cite{siegmund15}.
This includes a catalog of 39 factors that may influence the results of program comprehension studies.
Many of them have parallels in our discussion.
However, we place greater focus on the considerations involved in the decisions made when designing an experiment.

Finally, von Mayrhauser and Vans \cite{vonmayrhauser95} and Storey \cite{storey05} emphasize the theoretical underpinnings of program comprehension research.
While this is obviously important and worthy, our work is focused on the more technical aspects of making the observations in the first place.

The following sections consider the code, task, metrics, and experimental subjects.
In each the pertinent considerations are listed first, and then the pitfalls.

\section{The Code}

Experiments on code comprehension necessarily start with code.
But finding suitable code is not easy.

\subsection{Considerations}

\subsubsection{Code Scope}

A central question regarding the code to use in a program comprehension study is how much code to use.
There is a wide spectrum of options: a short snippet of a few lines, a method, a complete class, a package, or a full system.

The main consideration in favor of a limited scope is in cases where such a scope corresponds to the focus of the study.
For example, when investigating the effect of the names of parameters on the understanding of a function, it is natural to use complete functions \cite{avidan17}.
If investigating control structures, focused snippets containing a single program element reduce confounding effects.
For example, this was done by Ajami et al.\ in a study that found differences in understanding loops that count up and loops that count down \cite{ajami19}.
An additional consideration is that a limited scope allows for a manageable experiment, for example not extending beyond a single hour of the experimental subject's time.
 
But if the focus is on understanding as it is done during real development, e.g.\ to fix a bug report, a large volume of code should be used.
Ideally, the whole system should be available, just as it would be in a real world setting.
This is important since understanding a full system is quite different from understanding a limited amount of code \cite{levy19}.
Brooks suggests that the difficulty in understanding large systems is due to their myriad possible states \cite{brooks87}.

In the past something that passes for a full system could involve relatively little code, thereby enabling practical experiments on ``complete systems''.
For example, in the mid 1980s Littman et al.\ used a 250-line, 14-subroutine Fortran program that maintains a database of personnel records.
Today such a volume more realistically represents a single class.

As a result, experiments often make compromises.
For example, a bug fixing task may skip the stage of locating the relevant code in the system, and focus only on the actual fix of the function in which the bug occurs.

The alternative is to conduct large scale experiments.
For example, Wilson et al.\ asked graduate students to implement change requests in programs comprising about 100 Kloc, 800 classes, and 500 files \cite{wilson19}.
Sj{\o}berg et al.\ suggest that realistic experiments should be based on hiring professional programmers for relatively long periods of days to months \cite{sjoberg02,sjoberg03}.
In such a setting subjects can work in a realistic environment, including having access to all the relevant code.
This is important because comprehension --- like development --- is an incremental process.
It takes time and accumulates, and short experiments cannot evaluate this.
At best they can focus on a well-defined single step.

Another alternative is to observe professionals during their work.
This approach was taken by von Mayrhauser and Vans, who analyzed maintenance sessions of professionals working on large-scale full systems.
For example, in \cite{vonmayrhauser96} they report in detail on a 2-hour session devoted to porting client programs to a new operating system, and in \cite{vonmayrhauser97} they report on two 2-hour sessions, one fixing a bug and the other searching for the location to insert new code.

\subsubsection{Code Difficulty}

Probably the most important characteristic of code used in an experiment is that the code be appropriate for the task and the subjects.
It should not to too easy and not too hard.

As an example of easy code consider Listing 1 in \cite{busjahn15}.
This listing comprises 22 lines of code.
It defines a class Vehicle with a constructor and a method, followed by a main function that creates a Vehicle object and calls the method.
The method increments the vehicle's speed, subject to not going over a maximal value.
So essentially all this code does is to increment an integer.
Whether this is a problem depends on its use in the experiment.
The original experiment was to use an eye tracker to follow the code reading order, so very simple code is a suitable base case.
But using such code in a comprehension study would probably actually measure the ability to find the one line that does something.

As an example of hard code consider Figure 3 of \cite{beniamini17}.
This is 11 lines long, comprising an initialized array and a function.
The array is a lookup table.
The function calculates the number of 1 bits in an input buffer by using the top and bottom halves of each byte as indexes to the lookup table and summing.
While short, this code is non-trivial due to the use of bitwise operations to manipulate array accesses.

Assessing whether code is of suitable difficulty is hard, because this issue interacts with the subjects.
For example, if subjects don't know about bitwise operations, code using such operations becomes impenetrable.
A similar problem occurs when understanding the code requires specific domain knowledge.
This needs to be checked in pilot studies and during subject recruitment.

\subsubsection{Code Source}

A final issue is whether to use real or synthetic code.
We typically want the code to be representative of code written by developers for real applications.
It is unrealistic to try to create a representative sample of ``real'' code.
But we can at least use real code.

Volumes of real code are now freely available in open source repositories.
A possible concern is whether this is also representative of proprietary code.
There are dissenting opinions on which approach produces better code (and by implication, also clearer code).
Proponents of open source cite Linus's law, and claim that open source is better due to being subject to review by multiple users \cite{raymond:cab}.
Alternatively, proprietary code has been claimed to be better because it is more managed in terms of testing and documentation.

One major concern with using open source code is that the code was written by people who know what it is for.
So the code may rely on implicit domain knowledge or reflect unknown assumptions and constraints.
If experimental subjects in code comprehension experiments lack this knowledge, they will be unable to understand the code.
A possible solution is to use code from utility libraries \cite{avidan17}, or to otherwise ensure that domain knowledge is not required.

If you write code for the experiments, this needs to be justified.
The justification is usually that specific code features are required for the study.
For example, Ajami et al.\ wrote code snippets with exactly the same functionality using different programming constructs, to investigate the effect of these constructs on understanding \cite{ajami19}.

\subsection{Pitfalls}

Even when all the considerations are taken into account, problems with the code can threaten the validity of the study.

\subsubsection{Misleading Code}

Perhaps the biggest problem is unintentionally misleading code.
If the code is misleading, subjects may make mistakes not because of the studied effect but because they were misled.
Gopstein et al.\ have identified 15 coding practices that may be misleading \cite{gopstein17}.
Examples include using an assignment as a value, using short-circuit logic for control flow (in \code{A$||$B}, if \code{A} is true \code{B} is not evaluated), or presenting literals in an unnatural encoding like octal.
Scalabrino et al.\ define a metric for the deceptiveness of code based on the discord between perceived comprehension and actual comprehension \cite{scalabrino:metrics}.

A major factor in misleading code appears to be names.
Arnaoudova et al.\ call this ``linguistic antipatterns'', e.g.\ when a variable's name does not match its type, or when its plurality does not match its use \cite{arnaoudova16}.
A simple example appears in Listing 4 of \cite{hofmeister19}.
This includes an array named \code{str}, and a line \code{int len = str.Length}.
But contrary to what you might expect, \code{str} is not a string.
Rather, ``str'' is an abbreviation for ``start'', and is used to denote the initial array of integers passed to a function that will change it.
To further confuse matters, the array that will be appended to \code{str} is called \code{end}, a name that may be more suitable for the final result.

A striking example of the effect of misleading names was given by Avidan and Feitelson \cite{avidan17}.
In a study about variable naming they used 6 real functions from utility libraries.
Each function was presented either as it was originally written, or with variable names changed to \code{a}, \code{b}, \code{c}, etc.\ in order of appearance.
The unexpected result was that in 3 of the functions there was no significant difference in the time to understand the different versions, and moreover, several subjects made mistakes --- and all the mistakes were in the versions with the original names.
The conclusion was that the names were misleading, to the degree of being worse than meaningless names like consecutive letters of the alphabet.
But presumably the original developers did not intentionally choose misleading names.
So the real threat is that names that look OK to you would turn out to be misleading to your experimental subjects.

A more insidious example comes from Soloway and Ehrlich's seminal paper on programming knowledge \cite{soloway84}.
The code samples shown in Figure 1 of that paper were meant to investigate the rule that ``a variable's name should reflect its function''.
This was done by writing code that calculates the maximum or the minimum of a set of input numbers.
The experimental subjects' task was to insert the correct relation symbol ($<$ or $>$) in the expression comparing the result so far with each new input.

But in the version calculating the minimum, the code used the name \code{max} instead of \code{min}.
This left the initialization to a large number instead of to 0 as the only clue that the code actually calculates the minimum; if you assume that the code calculates the maximum, as the variable name suggests, such an initialization would be erroneous.
In other words, the experiment did not presented its subjects with a situation in which the name does not reflect its function --- it presented them with a downright \emph{contradiction}.
And this contradiction pitted a central variable name against a not so prominent initialization.
Subjects would have to be especially diligent to get this right.

\subsubsection{Recognized Code}

The opposite of misleading code is easily recognized code.
This may occur if textbook examples are used, e.g.\ a well-known sorting algorithm.
Using such code may end up measuring how well-versed subjects are in the cannon of programming examples.
Modifying such code risks turning it into misleading code, because subjects who recognize it expect the conventional unaltered functionality.
For example, altering the initialization or termination of a canonical \code{for} loop leads to a large increase in the errors made in interpreting what it does \cite{ajami19}.

A special case is when the code allows the answer to be guessed.
For example, \cite{sharif10} studied name recall using multiple-choice questions with the following options: \code{fill\_pathname}, \code{full\_mathname}, \code{full\_pathname}, and \code{full\_pathnum}.
Only one of these makes sense as a variable name.

\subsubsection{Problems with Structure, Style, or Dead Code}

Code used in experiments should be realistic, in the sense that it could have been written in the context of a real project.
Code written specifically for experiments sometimes violates this requirement.
For example, it should not include parts that do not make sense, especially if such unnecessary additions may give the experiment away.

An example can be found in \cite{schankin18}, which uses a class that converts variable names in under\_score style to camelCase style.
The class contains two helper functions, to lowercase the first letter of a word and to capitalize it.
The point of the experiment is to notice that \code{lowercase} is called instead of \code{capitalize}, which is an error.
But in fact there is no reason for the \code{lowercase} function to exist at all, because the class as presented supports only one-way conversion.
And in the original (erroneous) code \code{capitalize} is never called, which may be a hint.

Another aspect is coding style.
Fashions change, and difference people write in different styles.
A mismatch between the writing style and the preferences (or experience) of the experimental subject may cause bias and a confounding effect.
Two things can be done to avoid such problems.
First, be consistent and use the same style throughout.
Second, match the style to the preferences of the subjects either when writing the code or when recruiting subjects.

\subsubsection{Badly Presented Code}

A potential problem in presenting code in the context of comprehension studies is what to do with comments and descriptive names.
For example, header comments and method names are specifically designed to allow readers to understand what a function does without reading its code.
Leaving them intact may therefore undermine an experiment where subjects are supposed to deduce just that.
But given that they normally exist, removing them creates an unnatural situation.
Thus this should only be done in experiments using short code segments, where the focus justifies using an unrealistic setting.
If a large body of code is used, names and comments should be retained.
Likewise, in experiments where the task is not to understand what the code does (e.g.\ a debugging task) there is no problem.

A more delicate issue is how to handle mathematical and logical expressions.
For example, should one rely on operator precedence, or use parentheses to clarify the order of evaluation?
Again, if this is not the issue being studied, the best approach is to make it as unobtrusive as possible.
Any effort that subjects spend on understanding expressions, and any mistakes they make, dilute the results that the experiment was designed to produce.
In practical terms, this means to make the expressions as simple and obvious as possible, including by using parentheses.

More generally, all aspects of code readability affect its comprehension.
If they are not the issue being studied (e.g.\ \cite{oman90}), they should be controlled.
When using real code it may be tempting to present the code as it was written.
But if the original code is not laid out properly, or uses an idiosyncratic style, this could introduce a confounding effect.
A good practice is to use an IDE's default indentation and syntax highlighting.
Methods within a class should be listed in calling order, meaning that called methods are placed after methods that call them \cite{geffen16}.

\subsubsection{Learning and Fatigue effects}

When multiple codes are used, The question arises in what order to display them.
Using the same order for all experimental subjects reduces variability and enhances comparisons.
However, such a consistent order may cause a confounding effect due to learning, fatigue, or dropouts.
For example, if unsuccessful subjects feel discouraged and drop out of the experiment, only the more successful subjects will reach the last questions \cite{ajami19}.
In other words, a difference in performance on different codes may be the result of their placement in the sequence, rather than a result of the differences we wish to study.
The common solution is to randomize the order.

\subsubsection{Variable Naming Side-Effects}

Variable names and function names are instrumental for comprehension, and in many cases they provide the main clues regarding what the code is about.
In the context of comprehension studies this may be undesirable, so the names have to be stripped of meaning.
Some of the ways that this has been done are problematic:
\begin{itemize}
    \item Simple options are using arbitrary strings (asdf, qetmji) or unrelated words (superman, purple).
    These are distracting, and may be useful only in relation to extreme research questions on reading or the possible detrimental effect of extremely bad (distracting) names.
    They should not be used if this is not the research issue.
    \item Another approach is to use obfuscation, e.g.\ by applying a simple letter-exchange cipher \cite{siegmund17}.
    This leads to names that are long and distracting non-words.
    For example, the function name `countSameCharsAtSamePosition' can change into `ecoamKayiEoaikAmKayiEckqmqca'.
    Long names that differ in just a couple of letters may become very hard to distinguish.
 
    Note too that obfuscation may deeply affect how subjects perform tasks.
    Variable names convey meaning, and thus enable a measure of top-down comprehension.
    Siegmund et al.\ therefore used obfuscated variable names to force subjects to use bottom-up comprehension based on the syntax \cite{siegmund14,siegmund17}.
    Such an effect may happen with problematic code also when this is not intentional.
\end{itemize}

Alternative better ways to obfuscate variable names are the following:
\begin{itemize}
    \item One option is to use words that represent the technical use of the variables, but do not reveal the intent --- exactly the opposite of what we usually try to do.
    For example, you could use \code{num1} and \code{num2} instead of \code{base} and \code{exponent} in a function that calculates a power \cite{siegmund14}.
    \item Perhaps the simplest and most straightforward approach is to just use consecutive letters of the alphabet in order of appearance \cite{avidan17,hofmeister19}.
    Note that this is different from using the first letter of the ``good'' name, as that may still convey information \cite{beniamini17}.
\end{itemize}

A related case is when names are abbreviated to see the effect of such abbreviations.
For example this can be done by concatenating the first 3 consonants in the name \cite{hofmeister19}.
However, this may lead to unnatural or misleading names, such as \code{str} for \code{start} or \code{rsl} for \code{result}.
It is better to use judgement rather than a mechanical approach, e.g.\ allowing \code{res} for \code{result} and the 4-letter \code{conc} for \code{concatenate}.

\subsubsection{Appropriateness for Task}

Importantly, the task subjects are required to perform and the code must be compatible.
For example, when studying whether indentation aids comprehension, one needs a task that depends on the block structure of the code.
Otherwise indentation is indeed not an important feature, and the results will show that it does not matter.
But this would be wrong, because maybe indentation does indeed matter for another task --- for example, one that is related to navigation and identification of code blocks.

For example, \cite{miara83} conducted a study on indentation using 102-line long code with a main and two functions, and a maximal nesting level of 3.
The result was that nesting had some effect, based on questions such as whether all variables were global, which require the definition of variables at the beginning of functions to be identified (the study is from 1983 and the code was written in Pascal).
Many years later \cite{bauer19} replicated this study, but the code used was 17-line single functions with a maximal nesting of 2, and the task was to anticipate what the program would print.
In this setting indentation was not found to be important, but maybe the reason was an inappropriate choice of code.

\section{The Task}

If we focus on comprehension per se, experiments on code comprehension are a sort of challenge-response game.
The experimenter challenges the subject to understand some code.
A subject that claims to have achieved such understanding must prove it by performing some task.
It is therefore vital that the task really reflect comprehension.
In a sense, the task defines what ``comprehension'' means.

\subsection{Considerations}

In real life program comprehension is rarely an end in itself.
Rather, comprehension is a prerequisite to perform some programming task, such as fixing a bug or adding a feature.
So experiments can use such tasks directly.

Alternatively one can consider comprehension itself.
In this case the main consideration in selecting the task to perform is that the task reflects the level of understanding in which you are interested.
Should the subjects just know the variables and data structures? Maybe the behavior at runtime? Or perhaps also the underlying algorithms?
The following subsections detail several such possible levels.
We name them using the straightforward dictionary meaning of different words.
However, this is not universal, and over the years these names have been used in various non-compatible ways.

\subsubsection{Reading Task (tokens and structure)}

The most basic level is just reading the code, so the task needs to assess the readability of code.
Note that readability is not the same as understanding \cite{buse10}.
``Reading'' is just the technical issue of recognizing tokens and structure.
However, the readability of code also has a strong influence on navigation in the code and on the findability of key elements in it \cite{oman90}.
Specific tasks that assess readability include
\begin{itemize}
    \item Find a certain word, e.g.\ the use of a variable.
    \item Identify nesting of constructs, e.g.\ the most deeply nested one.
    \item Verify whether two expressions have the same syntactic structure.
\end{itemize}

\subsubsection{Parsing Task (understand syntax)}

The next level up is to be able to parse the code.
This shows that you are able to understand the syntax: what are leagal expressions, and what their relations may be.
Example tasks can include
\begin{itemize}
    \item Find the type of a variable (in a typed language).
    \item Find a syntax error in a function.
    \item At a larger scale, draw a UML class diagram of a project.
\end{itemize}

\subsubsection{Interpretation Task (local semantics)}

While parsing requires understanding the structure of the code, interpretation requires understanding the semantics of the individual instructions.
Tasks which reflect the ability to interpret code include
\begin{itemize}
    \item Find what the code prints for a certain input.
    This can be done by simulating the execution one instruction at a time, much like an interpreter would.
    \item Draw a UML sequence diagram.
    To do so all you need is to understand which functions call each other.
\end{itemize}

\subsubsection{Comprehension Task (global semantics)}

Comprehension is understanding the underlying concepts of the code: to be able to explain its functionality in abstract terms.
Specific tasks which reflect comprehension are:
\begin{itemize}
    \item After reading and understanding the code, answer a question about the expected output for a given input (without seeing the code again).
    \item Describe the flow of the code.
    Or more generally, any code summarization task.
    \item Add documentation to the code, for example header comments for functions.
    Or more specifically: suggest a suitable meaningful name for a function.
    \item Write a test suite for a function.
    Specifically show understanding of semantic corner cases.
    \item Explain the limitations of a function or API --- when should it be used, and when can't it be used.
    \item Articulate the contract of a function or API: what are the preconditions and postconditions \cite{meyer92}. 
\end{itemize}
Note that the difference between semantics and syntax is real.
fMRI studies shows that comprehension tasks activate different parts of the brain than syntax-related tasks --- parts related to working memory, attention, and language processing \cite{siegmund14}.

\subsubsection{Use Task (black-box)}

Black-box is a special case of comprehension, where we are interested in using the code as opposed to understanding how it works (white-box) \cite{levy19}.
This is very common and important in real life, and forms the basis for modularity, encapsulation, and information hiding \cite{parnas72,parnas85b}.
But it is rather uncommon and hard to use in comprehension experiments.
The simplest way to exhibit black-box knowledge about an API is to use it, that is to write some code that calls the API functions.
In addition, one can ask about various attributes of the API:
\begin{itemize}
    \item Details about parameters of API functions.
    \item Connections between functions, e.g.\ if one must be called before another is called.
    \item Documented preconditions or constraints.
\end{itemize}

Note that black-box understanding is actually disconnected from the code itself --- this is the essence of information hiding.
It is based on documentation.
But \emph{generating} such knowledge requires deeper comprehension, as noted above.

\subsubsection{Correction Task (white-box)}

Of the different types of maintenance \cite{lientz78}, corrective maintenance (fixing bugs) is the one most often used to test understanding.
But not all bugs reflect the same level of understanding.
One needs to distinguish technical bug fixing (e.g.\ finding and correcting a null pointer dereference \cite{levy19} or a syntax error)
from a semantic error (such as calling the wrong helper function \cite{schankin18} or using a wrong index into an array \cite{hofmeister19}).
Finding technical errors is more at the level of interpretation than comprehension --- it can be done by scanning the code superficially without any deep understanding of the whole.
Syntax errors may be irrelevant, as they should be caught by the compiler.
Only semantic errors reflect real understanding, similar to comprehension.

An important question is exactly what bugs to inject.
Two classifications were suggested by Basili and Selby \cite{basili87}.
The first is a distinction between errors of omission and errors of commission.
This is an important distinction, because with commission the subjects can see the error, but for omission they need to notice that something is missing --- which depends on a preconception of what the code is trying to do.
The second classification lists six types: initialization, control, computation, interface, data, and cosmetic.
These classifications were used for example in \cite{juristo12,jbara14b}.

\subsubsection{Extension or Modification Task (large scale white-box)}

Most of the tasks outlined above are suitable for short code snippets, a function, or perhaps a class.
Some of them, e.g.\ correction tasks, can also apply to larger software systems.
Extension and modification of software can also be done on a single function, but usually the minimal relevant scope is a class, and the common scope in real-life situations is a module or a complete system.

A few examples are given by Wilson et al.\ who use large-scale projects of 78 and 100 KLoC to study adding new features \cite{wilson19}.
This enables them to study not only the change itself, but also the process of finding where in the code the change must be made.
If the task asks only to change a given function, it misses the steps of zeroing in on the correct location to make the change, and the evaluation of the impact that the change may have on other parts of the system \cite{rajlich02}.

\subsubsection{Design-Related Task (abstraction)}

Understanding a system is not the same as understanding a single module or a smaller piece of code \cite{levy19}.
When understanding a system the focus is on understanding the structure, namely the system's components and how they interact with each other.
A deeper level of understanding is to understand why it is structured like this, that is, to understand the rationale for the design decisions that were taken during development.

The result of the design process is an architecture.
Hence understanding the design is understanding the architecture.
This can be expressed, for example, as defined by the 4+1 viewes suggested by Kruchten \cite{kruchten95}.
In an experiment the task can then be to draw a conceptual diagram showing relations between entities.
Note, however, that in a real-life setting comprehending a system is a continuous process, and that each task adds to this understanding in an incremental manner \cite{vonmayrhauser95}.
In all likelihood such a process cannot be replicated in an experiment.

\subsubsection{Recall Task}

A rather different type of task is to read the code, understand it, and then try to recall it from memory.
Obviously this is limited to reasonably short codes, e.g.\ up to 20--30 lines long.
The motivation for this task is the seminal work of Simon and Chase, which showed that expert chess players can easily memorize meaningful chess positions, but are not good at memorizing random placements of chess pieces \cite{simon73}.
Hence the memorization interacts with identification of meaning.

Shneiderman conducted an experiment based on this approach more than 40 years ago \cite{shneiderman77}, concluding that better recall indeed correlates with better comprehension (as measured by the ability to make modifications to the code).
McKeithen et al.\ also showed that expert programmers are better able to recall semantically meaningful program code \cite{mckeithen81}.
However, this type of task is rather far removed from what programmers actually do, and perhaps for this reason does not seem to be popular.

\subsubsection{Using Multiple Tasks}

The previous subsections indicate that different tasks actually reflect difference aspects of understanding.
To get a fuller picture it may therefore be advisable so use multiple different tasks in the same study.
With multiple tasks one can also obtain a more nuanced measure of success, as reflected by the fraction of the tasks that were completed correctly.

\subsubsection{Imposing Time Limits}

There are basically two approaches to measuring performance: how much you can achieve in a given time, or how long it takes to perform a given task \cite{bergersen14}.
Most experiments on comprehension measure time for a task.
This is also closer to normal working conditions.
However, placing a generous time limit may be advisable to exclude subjects who experience difficulties for some reason.

\subsection{Pitfalls}

\subsubsection{Levels of Understanding}

The main problem with selecting a task that reflects program understanding is the classic construct validity issue: are you measuring what you set out to measure?
In particular, does your task actually measure understanding at the level you are interested in?

While this is mainly a matter of definitions and word semantics, it is indeed important to agree on concepts and their definitions.
For example, consider Buse and Weimer's ``A metric for software readability'' \cite{buse08}, which --- very naturally, given its title --- is often cited as a reference on readability.
But in the reported experiments, subjects were told to score code snippets ``based on [your] estimation of readability'', where ``readability is [your] judgment about how easy a block of code is to understand''.
According to the definitions given above readability is just one of the factors that make code easy to understand.
So in our terminology Buse and Weimer's experiment is closer to measuring comprehensibility than readability.

Another problem is how you actually confirm that comprehension has indeed been achieved.
Scalabrino et al., for example, distinguish between \emph{perceived} understanding, where subjects just declare that they think they have understood a method, and \emph{actual} understanding, where they correctly answer several verification questions \cite{scalabrino:metrics}.
The questions they suggested were about the meaning of a variable name, or the purpose of calling a certain function.
It is debatable whether such questions indeed reflect a full understanding of the code.

\subsubsection{The Danger of Shortcuts}

As noted above, participating in a comprehension experiment is a challenge.
But experimental subjects may be lazy \cite{roehm12,levy19}.
They may prefer to use an ``as-needed'' program comprehension strategy as an alternative to a ``systematic'' strategy leading to full understanding \cite{littman87}.
Indeed, given a specific task, one can make do with comprehending only what is needed for this task \cite{vonmayrhauser98}

When designing an experiment it is therefore of paramount importance to avoid tasks where the brunt of the work can be avoided.
This is a significant threat to the premise that comprehension is a prerequisite for testing, debugging, and maintenance.
Examples include cases where tasks can be done mechanically without understanding.
For example, in bug fixing, finding a syntax error or finding a null pointer reference can be done without understanding what the function does.
In code modification, a simple refactor like function extract can be done without understanding what the function does.

\subsubsection{Confounding Explanations}

In controlled experiments one needs a control: a base-level treatment with which to compare the performance on the other treatments.
This is what gives controlled experiments their explanatory power (which is why it is regrettable that there is such a limited use of controlled experiments \cite{sjoberg05}).

To provide explanatory power the task has to be crisp in the sense that it strongly supports a certain interpretation.
Not all tasks have this property.
For example, the fill in the blanks task used by Soloway and Ehrlich is not crisp, because the variable name they used is misleading (as described above) \cite{soloway84}.
Thus a failure to answer correctly may not be due to a problem with the conceptual model (what the experiment was supposed to check), but simply due to falling in the trap of the misleading name.
Likewise, failure in recalling code verbatim from memory may identify totally wrong code or code that does not abide by conventions, not necessarily hard to understand code.

\subsubsection{The Working Environment}

A potentially important confounding factor is the working environment in which subjects perform their task.
Certain environments may include facilities that support the task and make it easier to complete.
If such an environment is provided, performing the task becomes easier.
However, this also depends on the subject knowing how to use the environment.
Subjects who do not know how to use the required feature (or don't know it exists) will be at a disadvantage.

The problem with features that support the task is that they may undermine the need to understand the code or affect the process of how it is understood.
In addition, if some subjects know how to use these features and others do not, this becomes a confounding factor that may interfere with the results.

A possible solution to this problem is to use a reduced environment, which does not include the features that may be used to help perform the task.
However, this is also problematic for subjects who are used to work in an environment that does include such support.

\section{The Metrics}

Rajlich and Cowan suggested that the dependent variables measured in comprehension studies should be the accuracy of the answers, the response time of accurate answers, and the response time of inaccurate answers \cite{rajlich97}.
Of these, the most commonly used is time to correct answer.
Accuracy is also often used, especially when it is easy to assess (for example, when the task is to predict what a given code will print).
The time to inaccurate answers is typically not used.
Recently, biophysical indicators (ranging from skin conductance through pupil size to fMRI brain activity patterns) have also been suggested as indicative of effort expended in code comprehension.

\subsection{Considerations}

The main consideration regarding metrics is that they be measurable.
This may interact with the task, as some tasks produce outcomes that are more measurable than others.

\subsubsection{Judging Accuracy}

If we consider comprehension experiments as a challenge-response game, the outcome of the game depends on the evaluation of the response.
If the response was correct, the experimental subject has met the challenge and ``wins''.
But how do we know whether the response was correct?
This obviously depends on the details of the task.

The easy cases are when the response is well-defined in advance, such as to identify what a given code will print (e.g.\ \cite{ajami19}).
In this case the answer can be checked automatically.
The only reservation is that inconsequential variations (e.g.\ an added space) should be ignored.
If multiple tasks are used, the fraction performed correctly can serve as a score.

In cases such as when the question is ``what does this code do'' or ``give this function a meaningful name'', one needs to prepare a capacity for judging the responses.
This should include
\begin{itemize}
    \item A key for judging correctness.
    For example, in an experiment based on comprehension of a program that created a histogram of word occurrences in a text, a third of the points were given for answering that the program counts word occurrences, a third for saying that it prints each unique word, and a third for noting that it prints the number of occurrences next to each word \cite{miara83}.
    \item Application of the key by multiple judges.
    \item A protocol for settling disputes, e.g.\ majority vote (2 of 3 judges) or conducting a joint discussion till reaching consensus.
\end{itemize}

\subsubsection{Reaction to Errors}

In those cases where a wrong answer can be detected automatically, e.g.\ when the experimental subject is required to find out what the code will print, one has to decide what to do if a wrong answer is given.
A common approach is to just go on with the experiment.
Alternatives include
\begin{itemize}
    \item Display a message indicating that a mistake has been made.
    But this may affect the rest of the experiment, either due to discouraging the subject, or due to facilitating a learning effect.
    \item In addition to indicating that a mistake was made, allow the subject to try again \cite{hofmeister19}.
    This raises the questions of how to measure time.
    Do you include the sum of all trials?
    Is it fair to compare this to the time taken by someone who did not try and fail?
    \item When it is expected that all subjects will succeed (which implies that correctness is not being measured), discard subjects who fail \cite{hofmeister19}.
    In other words, failure is used as an exclusion criterion.
\end{itemize}

\subsubsection{Dimensions of Performance}

If both time and accuracy (correctness) are measured, the question is whether to report them separately or to combine them in some way.
Combining the two metrics simplifies the analysis by making it one-dimensional.
But this is justified only if they indeed reflect the same underlying concept.

Bergersen et al.\ suggest a crude categorical classification scheme which combines time and correctness \cite{bergersen11b}.
In its simplest form, this scheme defines 3 levels of accomplishment:
\begin{enumerate}
    \item Incorrect answer.
    \item Correct answer, time above the median.
    \item Correct answer, time below the median.
\end{enumerate}
If the task is made up of multiple stages, the levels first reflect the number of stages completed successfully, and if all where, the time range in which this was achieved.

Beniamini et al.\ suggest a continuous version, where accomplishment is defined to be the quotient of the correctness score divided by the time \cite{beniamini17}.
This can be interpreted as the ``rate of answering correctly''.
Incorrect answers are naturally included with a rate of 0.
Scalabrino suggest a similar formula, but use the time saved relative to the subject who took the longest to answer \cite{scalabrino:metrics}.
This has the disadvantage that outliers may distort the results of others.

A related question is what is the significance of time to incorrect answer?
The most common approach is to ignore this data.
A possible alternative is to interpret such data as instances of censoring: we know that the subject spent this much time and did not arrive at a correct answer, therefore the time needed for a correct answer would be longer.

\subsubsection{Eye Tracking and Biometrics}

In recent years there is an increasing use of eye trackers in the context of code comprehension studies \cite{shaffer15,sharafi15,obaidellah18,bednarik20,sharafi20}.
eye trackers enable an identification and quantification of how the experimental subjects focus on different parts of the code, and also a recording of the gaze scan path: the order in which they go over the code.

Eye tracking is especially useful to identify what the experimental subjects are interested in.
An example is given by Jbara and Feitelson \cite{jbara17}.
This study used eye tracking to quantify the amount of time spent looking at successive repetitions of the same basic structure.
The results showed that the first instances get more attention, and were used to create quantitative models of how attention decreases with instance serial number.

In addition to eye tracking, various biophysical indicators for effort have been used in relation to software engineering research \cite{fritz14}.
Perhaps the most prominent is fMRI \cite{siegmund14,siegmund17,floyd17,ivanova20}.
For example, this has enabled the distinction between brain activity patterns when performing syntactic vs.\ semantic tasks \cite{siegmund14}.

\subsection{Pitfalls}

\subsubsection{Confounding Effects}

Measurements are always subject to the danger of confounding effects.
Many of the pitfalls noted in the previous sections may come into play when we measure the time or accuracy of code comprehension, and lead to unreliable results --- namely results which do not reflect the intended aspects of code comprehension.

One straightforward effect is getting used to the experimental setting.
It is apparently not uncommon that the first task or two in a sequence take longer, as the subjects learn what exactly is required of them (e.g.\ Figure 5 in \cite{ajami19}).
It may therefore be better to discard the first such result(s), or use them to evaluate the participants.

A special case is the relation between time and correctness.
Errors by definition reflect misunderstandings.
The question is whether this is due to difficulty of the code structure or to misleading beacons.
Evidence that time and correctness may actually reflect different concepts is given by Ajami et al.\ \cite{ajami19}.
This study included a comparison of understanding a canonical \code{for} loop (\code{for (i=0; i$<$n; i++)}) with variations in which the initialization, termination condition, or step are varied.
The results were that loops counting down took a bit longer, while loops with abnormal initialization or termination caused more errors.
The suggested interpretation was that time reflects difficulty, and the error rate reflects a ``surprise factor'', namely whether the code deviates from expectations.
Thus if the code contains misleading elements it may be ill-advised to combine time and correctness scores.

\subsubsection{Measurement Technicalities}

The understanding of short code snippets may take a short time measured in seconds.
Thus inaccuracies in the times of beginning and ending the measurement may have an effect.
The beginning is typically when the code is first presented, and does not pose a problem.
But the ending time may be ambiguous in the sense that it may or may not also include the time to report the answer.
Hofmeister et al.\ explicitly use a two-step system \cite{hofmeister19}.
Their experimental platform first requires subjects to indicate that they have achieved comprehension.
It then stops the clock and freezes the code display, and only then opens a window where the subjects can enter their answer.
Wilson et al.\ use an Eclipse plugin which measures the time spent performing different actions to differentiate between time spent comprehending and time spent coding \cite{wilson19}.

An alternative to using eye trackers is to employ an experimental platform which displays the code via a ``letterbox slit''.
With this mechanism most of the screen is hidden from view, and only the lines in the slit are visible \cite{jansen03,hofmeister19}.
The slit may be moved up and down using the arrow keys.
This has the significant advantage of enabling online experiments over the Internet, instead of requiring subjects to come physically to the lab where the eye tracking device is set up.
However, it is an unnatural setting, which might impair performance.
In particular, subjects cannot use peripheral vision to observe the structure of the code and navigate directly to different locations.

\subsubsection{Premature Theorizing}

The ultimate goal of research on program comprehension is to formalize theories on the cognitive processes which underlie comprehension (e.g.\ \cite{brooksr83,vonmayrhauser95,storey05,siegmund16}).
These are then expected to inform and facilitate the design of better tools and methodologies for software development.
However, not every measurement should lead directly to a cognitive theory.
We need to collect a lot of data first.
In particular we need multiple replications of existing research, which is the way to increase our confidence in the results, and to better define their limitations and illuminate their nuances.

\section{The Experimental Subjects}

Different people exhibit different levels of performance in all human endeavors.
Many factors influence how people perform in code development.
Three important high-level factors that affect performance are \cite{campbell93}:
\begin{itemize}
\item \textbf{Knowledge} --- what a developer knows, e.g. the syntax of a programming language;
\item \textbf{Skill} --- the developer's aptitude in applying his or her knowledge in a given situation, and the degree to which this is done automatically or requires effort; and
\item \textbf{Motivation} --- how much the developer actually wants to develop, which affects the effort invested in applying the skill. 
\end{itemize}
The same factors also influence how people perform in code comprehension.
And as people may have different knowledge, different skills, and different levels of motivation, their performance will differ too.

\subsection{Considerations}

Previous sections were about technical aspects of program comprehension studies.
Despite the various problems and complications that were discussed, these are things that are relatively easy to control.
The biggest problem is the human element, namely the experimental subjects.
Helpfully, there have been several reviews and guides on this matter (e.g.\ \cite{ko15}).

\subsubsection{Variability and Representativeness}

Variability among humans is a huge confounding factor, which is hard to assess and control \cite{bergersen12,curtis14}.
Large individual differences have been reported in various empirical studies, e.g.\ \cite{sackman68,curtis81,prechelt99}.
This has two main implications.
First, studies have to contend with large variability, and use large enough samples and appropriate statistical methods.
Second, it is important to check whether the variability correlates with demographic variables, and assess whether this affects the external validity of the experiment.
Most of the following subsections are elaborations of this consideration.

\subsubsection{Using Students}

Most studies on software engineering, including those focused on comprehension, loosely target ``professional developers''.
That raises the question of whether performing experiments with students as subjects is appropriate \cite{falessi18}.
Feitelson lists the following potential problems with students \cite{feitelson:stud}.:
\begin{itemize}
    \item They may not have fully ingested what they had learned, or hold misconceptions regarding what they have learned \cite{kaczmarczyk10}.
    \item They may not know of commonly used tools or use them ineffectively.
    \item They lack practical experience, which makes it harder for them to find and focus on the heart of the issue.
    \item Their academic orientation may be misaligned with the needs in industry.
\end{itemize}
On the other hand students may be more consistent in following instructions, rather than trying to cut to the core in whatever way (including violating the experimental protocol).

In addition, the dichotomy pitting ``students'' against ``professionals'' is overly simplistic.
Students may have had professional experience in their past or work in parallel with their studies.
Graduating students are very close to novice professionals.
And even if real differences exist, experiments using students can still be useful for making scientific progress \cite{tichy00,basili07}, e.g.\ by helping to focus the research and by debugging experimental procedures.

\subsubsection{Work Experience}

An apparently better alternative to avoiding ``students'' is to require work experience.
Preferring subjects with a certain minimal level of work experience is an easy and accessible inclusion criterion.
A suitable threshold may be as low as 3--5 years of professional experience.
The reason is that, beyond a certain level, achievements tend to ``flatten out'' unless one puts in special effort to improve \cite{newell81,heathcote00,ericsson93}.

In order to create a meaningful distinction between novices and experienced programmers, Feitelson et al.\ suggest to define three groups \cite{feitelson:names}:
\begin{itemize}
    \item Novice students without significant programming experience outside their studies, e.g.\ in the second year of their undergraduate studies and with at most 2 years of programming experience.
    \item Professionals, with at least 5 years of programming for a living beyond any programming done during their studies.
    \item all those falling in between the above two groups.
    These are excluded from the analysis, to sharpen the distinction between students/novices and professionals.
\end{itemize}

However, one should note that experience does not necessarily imply high performance \cite{sonnentag06}.
And knowledge may be more important for skill than experience \cite{bergersen11}.
Falessi et al.\ stress the need to consider not only duration of experience, but that the experience is relevant and recent \cite{falessi18}.
Salman et al.\ note that when investigating new technology, even experienced professionals are actually like novices, because by definition they do not yet have experience with it \cite{salman15}.

\subsubsection{Proficiency and Skill}

What we are really interested in is the subjects' expertise.
Dreyfus and Dreyfus identified the following possible levels \cite{dreyfus80}:
\begin{enumerate}
    \item Novice: knows to apply learned rules to basic situations
    \item Competent: recognizes and uses recurring patterns based on experience
    \item Proficient: prioritizes based on holistic view of the situation
    \item Expert: experienced enough to do the above intuitively and automatically
\end{enumerate}

The preferred alternative to tagging subjects as students or professionals or to counting years of experience is to explicitly evaluate their proficiency \emph{in isolation} from the experimental framework.
However, distinguishing between the above levels is hard to do.
The suggested alternative is then to try to assess skill.
Bergersen et al.\ have suggested a testing regime that can take up to two days \cite{bergersen14}.
A possible alternative is to make do with self assessment of skill \cite{siegmund14b}.


\subsubsection{Effect of Demographics}

A recurring theme when considering experimental subjects is whether demographic variables, mainly sex and age, may explain some of the variability.
Some studies have reported observed differences between men and women \cite{lawrie06,sharafi12}.
Others have found no such differences \cite{ajami19,feitelson:names}.
At present it seems that the differences, when and if they exist, are not major, but this deserves further study.

\subsubsection{Ethics}

Carver et al.\ point out that the considerations involved in performing quality research are not the whole story.
If using student subjects, one needs to remember that the students are there for an education, and participating in an experiment can affect this education \cite{carver10}.
It is up to the researchers to ensure that this effect is for the good.
More generally, care should be taken to ensure non-coercing participation and to limit stress.

\subsection{Pitfalls}

\subsubsection{Dimensions of Knowledge}

Developers often have different levels of knowledge in different pertinent dimensions of knowledge.
Therefore any uni-dimensional classification into ``experts'' and ``novices'' is compromised.

The most important distinction is between the technical dimension and the domain dimension \cite{shaft98}.
The technical dimension involves knowledge about the programming language, the development environment, the process workflows, etc.
Domain knowledge is about the background of the application: what exactly it is supposed to do, why, and how it fits into the bigger picture.
These two dimensions are required in different amounts for different tasks.
For example, technical knowledge is sufficient for fixing a technical bug like a null pointer reference, but domain knowledge is crucial for providing suitable context in a code summarization task.

Indeed, it is important to note that knowledge dimensions may interact with tasks.
According to von Mayrhauser and Vans, adaptive maintenance tasks require much more domain knowledge than program knowledge \cite{vonmayrhauser98}.
Corrective maintenance and the development of new features, on the other hand, require much more program knowledge than domain knowledge.

\subsubsection{Definition of ``Experts'' and ``Novices''}

Many studies claim to make a distinction between experts and novices.
But the definitions used differ considerably, making any comparison between different studies suspect.
For example, many use graduate students, or even students towards the end of their first degree, as ``experts''.
This might be true relative to freshmen in their first year, but does not reflect experience gained in a few years on the job.

\subsubsection{Subjects Unsuited for the Study}

A potentially significant problem may be the failure to exclude subjects who are unsuited for the task done in the experiment.
Subjects may not be well-versed in the programming language used.
This should be verified as part of the initial demographic screening.
They may lack knowledge needed to perform well in the study, e.g.\ knowledge about certain technologies.
Carver et al.\ suggest that this last deficiency can be corrected quickly by first observing someone else perform the experiment while using the required technology \cite{carver03b}.

Inexperienced subjects should not be used when the research involves not just basic or technical knowledge but performance honed by practice.
For example, naming or documentation practices may change after you have first-hand experience suffering from the practices of others.
So names given by experienced developers may differ from those given by inexperienced students \cite{feitelson:names}.

Last, subjects repeating the experiment should most probably also be excluded.
But if no identifying information is collected, this has to rely on subject self reporting.
To enable such reporting, a question needs to be included in the demographic screening.

\subsubsection{Unmotivated Subjects}

The performance on any work task may be affected by motivation, interest, and mood \cite{graziotin18}.
And motivation when participating in experiments may not be same as in real work.
Hofmeister et al.\ therefore suggest to exclude subjects who report that they were not working conscientiously in the final debriefing \cite{hofmeister19}.

\section{Conclusions}

Many of the points made in the previous sections may seem obvious.
But the literature is rife with examples of good research papers that did not take some of these considerations into account or failed on some pitfall.

The purpose of this checklist is not to dictate the ``right'' way to do research.
Its purpose is to raise awareness to the myriad considerations that are involved in experiments on program comprehension, and especially to the side effects that methodological decisions may have.
Such awareness is needed mainly to increase the volume of discussion of methodological issues, including methodological differences.
Awareness of differences is important for better understanding of how the results of different studies can be compared to each other, and how they complement each other.
This, together with multiple divergent replications of previous work, is the path to a deeper understanding of how code is understood.

%

\bibliographystyle{myabbrv}
\bibliography{abbrv,se,misc}

\begin{thebibliography}{10}\itemsep 0pt
\newcommand{\enquote}[1]{``#1''}
\providecommand{\url}[1]{\texttt{#1}}
\providecommand{\urlprefix}{URL }
\expandafter\ifx\csname urlstyle\endcsname\relax
  \providecommand{\doi}[1]{\textsf{DOI:\discretionary{}{}{}#1}}\else
  \providecommand{\doi}{\textsf{DOI:\discretionary{}{}{}\begingroup
  \urlstyle{rm}\Url}}\fi

\bibitem{ajami19}
S.~Ajami, Y.~Woodbridge, and D.~G. Feitelson, \enquote{\textsl{Syntax,
  predicates, idioms --- what really affects code complexity?}}
  \textit{Empirical Softw.\ Eng.} \textbf{24(1)}, pp.\ 287--328, Feb 2019,
  \doi{10.1007/s10664-018-9628-3}.

\bibitem{arnaoudova16}
V.~Arnaoudova, M.~Di~Penta, and G.~Antoniol, \enquote{\textsl{Linguistic
  antipatterns: What they are and how developers perceive them}}.
  \textit{Empirical Softw.\ Eng.} \textbf{21(1)}, pp.\ 104--158, Feb 2016,
  \doi{10.1007/s10664-014-9350-8}.

\bibitem{avidan17}
E.~Avidan and D.~G. Feitelson, \enquote{\textsl{Effects of variable names on
  comprehension: An empirical study}}. In 25th \textit{Intl.\ Conf.\ Program
  Comprehension}, pp.\ 55--65, May 2017, \doi{10.1109/ICPC.2017.27}.

\bibitem{basili87}
V.~R. Basili and R.~W. Selby, \enquote{\textsl{Comparing the effectiveness of
  software testing strategies}}. \textit{IEEE Trans.\ Softw.\ Eng.}
  \textbf{SE-13(12)}, pp.\ 1278--1296, Dec 1987, \doi{10.1109/TSE.1987.232881}.

\bibitem{basili86}
V.~R. Basili, R.~W. Selby, and D.~H. Hutchens, \enquote{\textsl{Experimentation
  in software engineering}}. \textit{IEEE Trans.\ Softw.\ Eng.}
  \textbf{SE-12(7)}, pp.\ 733--743, Jul 1986, \doi{10.1109/TSE.1986.6312975}.

\bibitem{basili07}
V.~R. Basili and M.~V. Zelkowitz, \enquote{\textsl{Empirical studies to build a
  science of computer science}}. \textit{Comm.\ ACM} \textbf{50(11)}, pp.\
  33--37, Nov 2007, \doi{10.1145/1297797.1297819}.

\bibitem{bauer19}
J.~Bauer, J.~Siegmund, N.~Peitek, J.~C. Hofmeister, and S.~Apel,
  \enquote{\textsl{Indentation: Simply a matter of style or support for program
  comprehension?}} In 27th \textit{Intl.\ Conf.\ Program Comprehension}, pp.\
  154--164, May 2019, \doi{10.1109/ICPC.2019.00033}.

\bibitem{bednarik20}
R.~Bednarik et~al., \enquote{\textsl{{EMIP}: The eye movements in programming
  dataset}}. \textit{Sci.\ Comput.\ Programming} \textbf{198}, art.\ no.\
  102520, Oct 2020, \doi{10.1016/j.scico.2020.102520}.

\bibitem{beniamini17}
G.~Beniamini, S.~Gingichashvili, A.~Klein~Orbach, and D.~G. Feitelson,
  \enquote{\textsl{Meaningful identifier names: The case of single-letter
  variables}}. In 25th \textit{Intl.\ Conf.\ Program Comprehension}, pp.\
  45--54, May 2017, \doi{10.1109/ICPC.2017.18}.

\bibitem{bergersen11}
G.~R. Bergersen and J.-E. Gustafsson, \enquote{\textsl{Programming skill,
  knowledge, and working memory among professional software developers from an
  investment theory perspective}}. \textit{J. Individual Differences}
  \textbf{32(4)}, pp.\ 201--209, Nov 2011, \doi{10.1027/1614-0001/a000052}.

\bibitem{bergersen11b}
G.~R. Bergersen, J.~E. Hannay, D.~I.~K. Sj{\o}berg, T.~Dyb{\aa}, and
  A.~Karahasanovi\'c, \enquote{\textsl{Inferring skill from tests of
  programming performance: Combining time and quality}}. In 5th \textit{Intl.\
  Symp.\ Empirical Softw.\ Eng.\ \& Measurement}, pp.\ 305--314, Sep 2011,
  \doi{10.1109/ESEM.2011.39}.

\bibitem{bergersen12}
G.~R. Bergersen and D.~I.~K. Sj{\o}berg, \enquote{\textsl{Evaluating methods
  and technologies in software engineering with respect to developer's skill
  level}}. In 16th \textit{Intl.\ Conf.\ Evaluation \& Assessment in Softw.\
  Eng.}, pp.\ 101--110, May 2012, \doi{10.1049/ic.2012.0013}.

\bibitem{bergersen14}
G.~R. Bergersen, D.~I.~K. Sj{\o}berg, and T.~Dyb{\aa},
  \enquote{\textsl{Construction and validation of an instrument for measuring
  programming skill}}. \textit{IEEE Trans.\ Softw.\ Eng.} \textbf{40(12)}, pp.\
  1163--1184, Dec 2014, \doi{10.1109/TSE.2014.2348997}.

\bibitem{brooks87}
F.~P. Brooks, Jr., \enquote{\textsl{No silver bullet: Essence and accidents of
  software engineering}}. \textit{Computer} \textbf{20(4)}, pp.\ 10--19, Apr
  1987, \doi{10.1109/MC.1987.1663532}.

\bibitem{brooksr83}
R.~Brooks, \enquote{\textsl{Towards a theory of the comprehension of computer
  programs}}. \textit{Intl.\ J.\ Man-Machine Studies} \textbf{18(6)}, pp.\
  543--554, Jun 1983, \doi{10.1016/S0020-7373(83)80031-5}.

\bibitem{brooksre80}
R.~E. Brooks, \enquote{\textsl{Studying programmer behavior experimentally: The
  problems of proper methodology}}. \textit{Comm.\ ACM} \textbf{23(4)}, pp.\
  207--213, Apr 1980, \doi{10.1145/358841.358847}.

\bibitem{buse08}
R.~P.~L. Buse and W.~R. Weimer, \enquote{\textsl{A metric for software
  readability}}. In \textit{Intl.\ Symp.\ Softw.\ Testing \& Analysis}, pp.\
  121--130, Jul 2008, \doi{10.1145/1390630.1390647}.

\bibitem{buse10}
R.~P.~L. Buse and W.~R. Weimer, \enquote{\textsl{Learning a metric for code
  readability}}. \textit{IEEE Trans.\ Softw.\ Eng.} \textbf{36(4)}, pp.\
  546--558, Jul/Aug 2010, \doi{10.1109/TSE.2009.70}.

\bibitem{busjahn15}
T.~Busjahn, R.~Bednarik, A.~Begel, M.~Crosby, J.~H. Paterson, C.~Schulte,
  B.~Sharif, and S.~Tamm, \enquote{\textsl{Eye movements in code reading:
  Relaxing the linear order}}. In 23rd \textit{Intl.\ Conf.\ Program
  Comprehension}, pp.\ 255--265, May 2015, \doi{10.1109/ICPC.2015.36}.

\bibitem{campbell93}
J.~P. Campbell, R.~A. McCloy, S.~H. Oppler, and C.~E. Sager, \enquote{\textsl{A
  theory of performance}}. In \textit{Personnel Selection in Organizations},
  N.~Schmitt, W.~C. Borman, and Associates (eds.), pp.\ 35--70, Jossey-Bass
  Pub., 1993.

\bibitem{carver03b}
J.~Carver, F.~Shull, and V.~Basili, \enquote{\textsl{Observational studies to
  accelerate process experience in classroom studies: An evaluation}}. In
  \textit{Intl.\ Symp.\ Empirical Softw.\ Eng.}, pp.\ 72--79, Sep 2003,
  \doi{10.1109/ISESE.2003.1237966}.

\bibitem{carver10}
J.~C. Carver, L.~Jaccheri, S.~Morasca, and F.~Shull, \enquote{\textsl{A
  checklist for integrating student empirical studies with research and
  teaching goals}}. \textit{Empirical Softw.\ Eng.} \textbf{15(1)}, pp.\
  35--59, Feb 2010, \doi{10.1007/s10664-009-9109-9}.

\bibitem{cherubini07}
M.~Cherubini, G.~Venolia, R.~DeLine, and A.~J. Ko, \enquote{\textsl{Let's go to
  the whiteboard: How and why software developers use drawings}}. In
  \textit{SIGCHI Conf.\ Human Factors in Comput.\ Syst.}, pp.\ 557--566, Apr
  2007, \doi{10.1145/1240624.1240714}.

\bibitem{curtis81}
B.~Curtis, \enquote{\textsl{Substantiating programmer variability}}.
  \textit{Proc.\ IEEE} \textbf{69(7)}, p.\ 846, Jul 1981,
  \doi{10.1109/PROC.1981.12088}.

\bibitem{curtis14}
B.~Curtis, \enquote{\textsl{A career spent wading through industry's empirical
  ooze}}. In 2nd \textit{Intl.\ Workshop Conducting Empirical Studies in
  Industry}, pp.\ 1--2, Jun 2014, \doi{10.1145/2593690.2593699}.

\bibitem{dreyfus80}
S.~E. Dreyfus and H.~L. Dreyfus, \textit{A Five-Stage Model of the Mental
  Activities Involved in Directed Skill Acquisition}. Tech. Rep.\ ORC-80-2,
  Operations Research Center, University of California, Berkeley, Feb 1980.

\bibitem{ericsson93}
K.~A. Ericsson, R.~T. Krampe, and C.~Tesch-R{\"o}mer, \enquote{\textsl{The role
  of deliberate practice in the acquisition of expert performance}}.
  \textit{Psychological Rev.} \textbf{100(3)}, pp.\ 363--406, Jul 1993,
  \doi{10.1037/0033-295X.100.3.363}.

\bibitem{falessi18}
D.~Falessi, N.~Juristo, C.~Wohlin, B.~Turhan, J.~{M\"{u}nch}, A.~Jedlitschka,
  and M.~Oivo, \enquote{\textsl{Empirical software engineering experts on the
  use of students and professionals in experiments}}. \textit{Empirical Softw.\
  Eng.} \textbf{23(1)}, pp.\ 452--489, Feb 2018,
  \doi{10.1007/s10664-017-9523-3}.

\bibitem{feitelson:stud}
D.~G. Feitelson, \enquote{\textsl{Using students as experimental subjects in
  software engineering research -- a review and discussion of the evidence}},
  Dec 2015. ArXiv:1512.08409 [cs.SE].

\bibitem{feitelson:names}
D.~G. Feitelson, A.~Mizrahi, N.~Noy, A.~Ben~Shabat, O.~Eliyahu, and R.~Sheffer,
  \enquote{\textsl{How developers choose names}}. \textit{IEEE Trans.\ Softw.\
  Eng.} \doi{10.1109/TSE.2020.2976920}. (early access).

\bibitem{floyd17}
B.~Floyd, T.~Santander, and W.~Weimer, \enquote{\textsl{Decoding the
  representation of code in the brain: An {fMRI} study of code review and
  expertise}}. In 39th \textit{Intl.\ Conf.\ Softw.\ Eng.}, pp.\ 175--186, May
  2017.

\bibitem{fritz14}
T.~Fritz, A.~Begel, S.~C. {M\"uller}, S.~Yigit-Elliott, and M.~{Z\"uger},
  \enquote{\textsl{Using psycho-physiological measures to assess task
  difficulty in software development}}. In 36th \textit{Intl.\ Conf.\ Softw.\
  Eng.}, pp.\ 402--413, May 2014, \doi{10.1145/2568225.2568266}.

\bibitem{geffen16}
Y.~Geffen and S.~Maoz, \enquote{\textsl{On method ordering}}. In 24th
  \textit{Intl.\ Conf.\ Program Comprehension}, May 2016,
  \doi{10.1109/ICPC.2016.7503711}.

\bibitem{gopstein17}
D.~Gopstein, J.~Iannacone, Y.~Yan, L.~DeLong, Y.~Zhuang, M.~K.-C. Yeh, and
  J.~Cappos, \enquote{\textsl{Understanding misunderstanding in source code}}.
  In 11th \textit{ESEC/FSE}, pp.\ 129--139, Aug 2017,
  \doi{10.1145/3106237.3106264}.

\bibitem{graziotin18}
D.~Graziotin, F.~Fagerholm, X.~Wang, and P.~Abrahamsson, \enquote{\textsl{What
  happens when software developers are (un)happy}}. \textit{J.\ Syst.\ \&
  Softw.} \textbf{140}, pp.\ 32--47, Jun 2018, \doi{10.1016/j.jss.2018.02.041}.

\bibitem{heathcote00}
A.~Heathcote, S.~Brown, and D.~J.~K. Mewhort, \enquote{\textsl{The power law
  repealed: The case for an exponential law of practice}}. \textit{Psychonomic
  Bulletin \& Review} \textbf{7(2)}, pp.\ 185--207, Jun 2000,
  \doi{10.3758/BF03212979}.

\bibitem{hofmeister19}
J.~C. Hofmeister, J.~Siegmund, and D.~V. Holt, \enquote{\textsl{Shorter
  identifier names take longer to comprehend}}. \textit{Empirical Softw.\ Eng.}
  \textbf{24(1)}, pp.\ 417--443, Feb 2019, \doi{10.1007/s10664-018-9621-x}.

\bibitem{ivanova20}
A.~A. Ivanova, S.~Srikant, Y.~Sueoka, H.~H. Kean, R.~Dhamala, U.-M. O'Reilly,
  <arina U.~Bers, and E.~Fedorenko, \enquote{\textsl{Comprehension of computer
  code relies primarily on domain-general executive brain regions}}.
  \textit{eLife} \textbf{9}, art.\ no.\ e58906, Dec 2020,
  \doi{10.7554/eLife.58906}.

\bibitem{jansen03}
A.~R. Jansen, A.~F. Blackwell, and K.~Marriott, \enquote{\textsl{A tool for
  tracking visual attention: The restricted focus viewer}}. \textit{Behavior
  Research Methods, Instruments, \& Comput.} \textbf{35(1)}, pp.\ 57--69, Feb
  2003, \doi{10.3758/BF03195497}.

\bibitem{jbara14b}
A.~Jbara and D.~G. Feitelson, \enquote{\textsl{On the effect of code regularity
  on comprehension}}. In 22nd \textit{Intl.\ Conf.\ Program Comprehension},
  pp.\ 189--200, Jun 2014, \doi{10.1145/2597008.2597140}.

\bibitem{jbara17}
A.~Jbara and D.~G. Feitelson, \enquote{\textsl{How programmers read regular
  code: A controlled experiment using eye tracking}}. \textit{Empirical Softw.\
  Eng.} \textbf{22(3)}, pp.\ 1440--1477, Jun 2017,
  \doi{10.1007/s10664-016-9477-x}.

\bibitem{juristo:exp}
N.~Juristo and A.~M. Moreno, \textit{Basics of Software Engineering
  Experimentation}. Kluwer, 2001.

\bibitem{juristo12}
N.~Juristo, S.~Vegas, M.~Solari, S.~Abrahao, and I.~Ramos,
  \enquote{\textsl{Comparing the effectiveness of equivalence partitioning,
  branch testing and code reading by stepwise abstraction applied by
  subjects}}. In 5th \textit{Intl.\ Conf.\ Software Testing, Verification, \&
  Validation}, pp.\ 330--339, Apr 2012, \doi{10.1109/ICST.2012.113}.

\bibitem{kaczmarczyk10}
L.~C. Kaczmarczyk, E.~R. Petrick, J.~P. East, and G.~L. Herman,
  \enquote{\textsl{Identifying student misconceptions of programming}}. In 41st
  \textit{SIGCSE Tech.\ Symp.\ Comput.\ Sci.\ Ed.}, pp.\ 107--111, Mar 2010,
  \doi{10.1145/1734263.1734299}.

\bibitem{ko15}
A.~J. Ko, T.~D. LaToza, and M.~M. Burnett, \enquote{\textsl{A practical guide
  to controlled experiments of software engineering tools with human
  participants}}. \textit{Empirical Softw.\ Eng.} \textbf{20(1)}, pp.\
  110--141, Feb 2015, \doi{10.1007/s10664-013-9279-3}.

\bibitem{kruchten95}
P.~Kruchten, \enquote{\textsl{The 4+1 view model of architecture}}.
  \textit{IEEE Softw.} \textbf{12(6)}, pp.\ 42--50, Nov 1995,
  \doi{10.1109/52.469759}.

\bibitem{lawrie06}
D.~Lawrie, C.~Morrell, H.~Field, and D.~Binkley, \enquote{\textsl{What's in a
  name? a study of identifiers}}. In 14th \textit{Intl.\ Conf.\ Program
  Comprehension}, pp.\ 3--12, Jun 2006, \doi{10.1109/ICPC.2006.51}.

\bibitem{levy19}
O.~Levy and D.~G. Feitelson, \enquote{\textsl{Understanding large-scale
  software -- a hierarchical view}}. In 27th \textit{Intl.\ Conf.\ Program
  Comprehension}, pp.\ 283--293, May 2019, \doi{10.1109/ICPC.2019.00047}.

\bibitem{lientz78}
B.~P. Lientz, E.~B. Swanson, and G.~E. Tompkins,
  \enquote{\textsl{Characteristics of application software maintenance}}.
  \textit{Comm.\ ACM} \textbf{21(6)}, pp.\ 466--471, Jun 1978,
  \doi{10.1145/359511.359522}.

\bibitem{littman87}
D.~C. Littman, J.~Pinto, S.~Letovsky, and E.~Soloway, \enquote{\textsl{Mental
  models and software maintenance}}. \textit{J.\ Syst.\ \& Softw.}
  \textbf{7(4)}, pp.\ 341--355, Dec 1987, \doi{10.1016/0164-1212(87)90033-1}.

\bibitem{martin:clean}
R.~C. Martin, \textit{Clean Code: A Handbook of Agile Software Craftmanship}.
  Prentice Hall, 2009.

\bibitem{mckeithen81}
K.~B. McKeithen, J.~S. Reitman, H.~H. Reuter, and S.~C. Hirtle,
  \enquote{\textsl{Knowledge organization and skill differences in computer
  programmers}}. \textit{Cognitive Pshchol.} \textbf{13(3)}, pp.\ 307--325, Jul
  1981, \doi{10.1016/0010-0285(81)90012-8}.

\bibitem{meyer92}
B.~Meyer, \enquote{\textsl{Applying ``design by contract''}}. \textit{Computer}
  \textbf{25(10)}, pp.\ 40--51, Oct 1992, \doi{10.1109/2.161279}.

\bibitem{miara83}
R.~J. Miara, J.~A. Musselman, J.~A. Navarro, and B.~Shneiderman,
  \enquote{\textsl{Program indentation and comprehensibility}}. \textit{Comm.\
  ACM} \textbf{26(11)}, pp.\ 851--867, Nov 1983, \doi{10.1145/182.358437}.

\bibitem{newell81}
A.~Newell and P.~S. Rosenbloom, \enquote{\textsl{Mechanisms of skill
  acquisition and the law of practice}}. In \textit{Cognitive Skills and Their
  Acquisition}, J.~R. Anderson (ed.), pp.\ 1--55, Lawrence Erlbaum Assoc.,
  1981.

\bibitem{obaidellah18}
U.~Obaidellah, M.~Al~Haek, and P.~C.-H. Cheng, \enquote{\textsl{A survey on the
  usage of eye-tracking in computer programming}}. \textit{ACM Comput.\ Surv.}
  \textbf{51(1)}, art.\ no.~5, Jan 2018, \doi{10.1145/3145904}.

\bibitem{oman90}
P.~W. Oman and C.~R. Cook, \enquote{\textsl{Typographic style is more than
  cosmetic}}. \textit{Comm.\ ACM} \textbf{33(5)}, pp.\ 506--520, May 1990,
  \doi{10.1145/78607.78611}.

\bibitem{parnas72}
D.~L. Parnas, \enquote{\textsl{On the criteria to be used in decomposing
  systems into modules}}. \textit{Comm.\ ACM} \textbf{15(12)}, pp.\ 1053--1058,
  Dec 1972.

\bibitem{parnas85b}
D.~L. Parnas, P.~C. Clements, and D.~M. Weiss, \enquote{\textsl{The modular
  structure of complex systems}}. \textit{IEEE Trans.\ Softw.\ Eng.}
  \textbf{SE-11(3)}, pp.\ 259--266, Mar 1985, \doi{10.1109/TSE.1985.232209}.

\bibitem{prechelt99}
L.~Prechelt, \enquote{\textsl{Comparing {Java} vs.\ {C/C++} efficiency
  differences to interpersonal differences}}. \textit{Comm.\ ACM}
  \textbf{42(10)}, pp.\ 109--112, Oct 1999, \doi{10.1145/317665.317683}.

\bibitem{rajlich97}
V.~Rajlich and G.~S. Cowan, \enquote{\textsl{Towards standard for experiments
  in program comprehension}}. In 5th \textit{IEEE Intl.\ Workshop Program
  Comprehension}, pp.\ 160--161, Mar 1997, \doi{10.1109/WPC.1997.601284}.

\bibitem{rajlich02}
V.~Rajlich and N.~Wilde, \enquote{\textsl{The role of concepts in program
  comprehension}}. In 10th \textit{IEEE Intl.\ Workshop Program Comprehension},
  pp.\ 271--278, Jun 2002, \doi{10.1109/WPC.2002.1021348}.

\bibitem{raymond:cab}
E.~S. Raymond, \enquote{\textsl{The cathedral and the bazaar}}. URL
  www.catb.org/\~{ }esr/writings/cathedral-bazaar/cathedral-bazaar, 2000.

\bibitem{roehm12}
T.~Roehm, R.~Tiarks, R.~Koschke, and W.~Maalej, \enquote{\textsl{How do
  professional developers comprehend software?}} In 34th \textit{Intl.\ Conf.\
  Softw.\ Eng.}, pp.\ 255--265, Jun 2012, \doi{10.1109/ICSE.2012.6227188}.

\bibitem{sackman68}
H.~Sackman, W.~J. Erikson, and E.~E. Grant, \enquote{\textsl{Exploratory
  experimental studies comparing online and offline programming performance}}.
  \textit{Comm.\ ACM} \textbf{11(1)}, pp.\ 3--11, Jan 1968,
  \doi{10.1145/362851.362858}.

\bibitem{salman15}
I.~Salman, A.~Tosun~Misirli, and N.~Juristo, \enquote{\textsl{Are students
  representative of professionals in software engineering experiments?}} In
  37th \textit{Intl.\ Conf.\ Softw.\ Eng.}, pp.\ 666--676, May 2015,
  \doi{10.1109/ICSE.2015.82}.

\bibitem{scalabrino:metrics}
S.~Scalabrino, G.~Bavota, C.~Vendome, M.~Linares-V\'squez, D.~Poshyvanyk, and
  R.~Oliveto, \enquote{\textsl{Automatically assessing code
  understandability}}. \textit{IEEE Trans.\ Softw.\ Eng.}
  \doi{10.1109/TSE.2019.2901468}. (early access).

\bibitem{schankin18}
A.~Schankin, A.~Berger, D.~V. Holt, J.~C. Hofmeister, T.~Riedel, and M.~Beigl,
  \enquote{\textsl{Descriptive compound identifier names improve source code
  comprehension}}. In 26th \textit{Intl.\ Conf.\ Program Comprehension}, pp.\
  31--40, May 2018, \doi{10.1145/3196321.3196332}.

\bibitem{shaffer15}
T.~R. Shaffer, J.~L. Wise, B.~M. Walters, S.~C. {M\"uller}, M.~Falcone, and
  B.~Sharif, \enquote{\textsl{{iTrace}: Enabling eye tracking on software
  artifacts within the {IDE} to support software engineering tasks}}. In
  \textit{ESEC/FSE}, pp.\ 954--957, Aug 2015, \doi{10.1145/2786805.2803188}.

\bibitem{shaft98}
T.~M. Shaft and I.~Vessey, \enquote{\textsl{The relevance of application domain
  knowledge: Characterizing the computer program comprehension process}}.
  \textit{J.\ Mgmt.\ Inf.\ Syst.} \textbf{15(1)}, pp.\ 51--78, 1998,
  \doi{10.1080/07421222.1998.11518196}.

\bibitem{sharafi20}
Z.~Sharafi, B.~Sharif, Y.-G. Gu\'eh\'eneuc, A.~Begel, R.~Bednarik, and
  M.~Crosby, \enquote{\textsl{A practical guide on conducting eye tracking
  studies in software engineering}}. \textit{Empirical Softw.\ Eng.}
  \textbf{25(5)}, pp.\ 3128--3174, Sep 2020, \doi{10.1007/s10664-020-09829-4}.

\bibitem{sharafi15}
Z.~Sharafi, Z.~Soh, and Y.-G. Gu\'eh\'eneuc, \enquote{\textsl{A systematic
  litareture review on the usage of eye-tracking in software engineering}}.
  \textit{Inf.\ \& Softw.\ Tech.} \textbf{67}, pp.\ 79--107, Nov 2015,
  \doi{10.1016/j.infsof.2015.06.008}.

\bibitem{sharafi12}
Z.~Sharafi, Z.~Soh, Y.-G. Gu\'eh\'eneuc, and G.~Antoniol,
  \enquote{\textsl{Women and men --- different but equal: On the impact of
  identifier style on source code reading}}. In 20th \textit{Intl.\ Conf.\
  Program Comprehension}, pp.\ 27--36, Jun 2012,
  \doi{10.1109/ICPC.2012.6240505}.

\bibitem{sharif10}
B.~Sharif and J.~I. Maletic, \enquote{\textsl{An eye tracking study on
  {camelCase} and {under\_score} identifier styles}}. In 18th \textit{Intl.\
  Conf.\ Program Comprehension}, pp.\ 196--205, Jun 2010,
  \doi{10.1109/ICPC.2010.41}.

\bibitem{shneiderman77}
B.~Shneiderman, \enquote{\textsl{Measuring computer program quality and
  comprehension}}. \textit{Intl.\ J.\ Man-Machine Studies} \textbf{9(4)}, pp.\
  465--478, Jul 1977, \doi{10.1016/S0020-7373(77)80014-X}.

\bibitem{shull:exp}
F.~Shull, J.~Singer, and D.~I.~K. Sj{\o}berg (eds.), \textit{Guide to Advanced
  Empirical Software Engineering}. Springer-Verlag, 2008.

\bibitem{siegmund16}
J.~Siegmund, \enquote{\textsl{Program comprehension: Past, present, and
  future}}. In 23rd \textit{Intl.\ Conf.\ Softw.\ Analysis, Evolution, \&
  Reengineering}, pp.\ 13--20, Mar 2016, \doi{10.1109/SANER.2016.35}.

\bibitem{siegmund14}
J.~Siegmund, C.~{K\"astner}, S.~Apel, C.~Parnin, A.~Bethmann, T.~Leich,
  G.~Saake, and A.~Brechmann, \enquote{\textsl{Understanding understanding
  source code with funcional magnetic resonance imaging}}. In 36th
  \textit{Intl.\ Conf.\ Softw.\ Eng.}, pp.\ 378--389, May 2014,
  \doi{10.1145/2568225.2568252}.

\bibitem{siegmund14b}
J.~Siegmund, C.~{K\"astner}, J.~Liebig, S.~Apel, and S.~Hanenberg,
  \enquote{\textsl{Measuring and modeling programming experience}}.
  \textit{Empirical Softw.\ Eng.} \textbf{19(5)}, pp.\ 1299--1334, Oct 2014,
  \doi{10.1007/s10664-013-9286-4}.

\bibitem{siegmund17}
J.~Siegmund, N.~Peitek, C.~Parnin, S.~Apel, J.~Hofmeister, C.~{K\"astner},
  A.~Begel, A.~Bethmann, and A.~Brechmann, \enquote{\textsl{Measuring neural
  efficiency of program comprehension}}. In 11th \textit{ESEC/FSE}, pp.\
  140--150, Sep 2017, \doi{10.1145/3106237.3106268}.

\bibitem{siegmund15}
J.~Siegmund and J.~Schumann, \enquote{\textsl{Confounding parameters on program
  comprehension: A literature survey}}. \textit{Empirical Softw.\ Eng.}
  \textbf{20(4)}, pp.\ 1159--1192, Aug 2015, \doi{10.1007/s10664-014-9318-8}.

\bibitem{simon73}
H.~A. Simon and W.~G. Chase, \enquote{\textsl{Skill in chess}}.
  \textit{American Scientist} \textbf{61(4)}, pp.\ 394--403, Jul-Aug 1973.

\bibitem{sjoberg02}
D.~I.~K. Sj{\o}berg, B.~Anda, E.~Arisholm, T.~Dyb{\aa}, M.~J{\o}rgensen,
  A.~Karahasanovic, E.~F. Koren, and M.~Vok\'ac, \enquote{\textsl{Conducting
  realistic experiments in software engineering}}. In \textit{Intl.\ Symp.\
  Empirical Softw.\ Eng.}, pp.\ 17--26, Oct 2002,
  \doi{10.1109/ISESE.2002.1166921}.

\bibitem{sjoberg03}
D.~I.~K. Sj{\o}berg, B.~Anda, E.~Arisholm, T.~Dyb{\aa}, M.~J{\o}rgensen,
  A.~Karahasanovi\'c, and M.~Vok\'a\v{c}, \enquote{\textsl{Challenges and
  recommendations when increasing the realism of controlled software
  engineering experiments}}. In \textit{Empirical Methods and Studies in
  Software Engineering: Experiences from {ESERNET}}, R.~Conradi and A.~I. Wang
  (eds.), pp.\ 24--38, Springer-Verlag, 2003,
  \doi{10.1007/978-3-540-45143-3\_3}. Lect.\ Notes Comput.\ Sci.\ vol.\ 2765.

\bibitem{sjoberg05}
D.~I.~K. Sj{\o}berg, J.~E. Hannay, O.~Hansen, V.~B. Kampenes,
  A.~Karahasanovi\'c, N.-K. Liborg, and A.~C. Rekdal, \enquote{\textsl{A survey
  of controlled experiments in software engineering}}. \textit{IEEE Trans.\
  Softw.\ Eng.} \textbf{31(9)}, pp.\ 733--753, Sep 2005,
  \doi{10.1109/TSE.2005.97}.

\bibitem{soloway84}
E.~Soloway and K.~Ehrlich, \enquote{\textsl{Empirical studies of programming
  knowledge}}. \textit{IEEE Trans.\ Softw.\ Eng.} \textbf{SE-10(5)}, pp.\
  595--609, Sep 1984, \doi{10.1109/TSE.1984.5010283}.

\bibitem{sonnentag06}
S.~Sonnentag, C.~Niessen, and J.~Volmer, \enquote{\textsl{Expertise in software
  design}}. In \textit{The Cambridge Handbook of Expertise and Expert
  Performance}, K.~A. Ericsson, N.~Charness, P.~J. Feltovich, and R.~R. Hoffman
  (eds.), pp.\ 373--387, Cambridge University Press, 2006.

\bibitem{storey05}
M.-A. Storey, \enquote{\textsl{Theories, methods and tools in program
  comprehension: Past, present and future}}. In 13th \textit{IEEE Intl.\
  Workshop Program Comprehension}, 2005.

\bibitem{tichy00}
W.~F. Tichy, \enquote{\textsl{Hints for reviewing empirical work in software
  engineering}}. \textit{Empirical Softw.\ Eng.} \textbf{5(4)}, pp.\ 309--312,
  Dec 2000, \doi{10.1023/A:1009844119158}.

\bibitem{vonmayrhauser95}
A.~von Mayrhauser and A.~M. Vans, \enquote{\textsl{Program comprehension during
  software maintenance and evolution}}. \textit{Computer} \textbf{28(8)}, pp.\
  44--55, Aug 1995, \doi{10.1109/2.402076}.

\bibitem{vonmayrhauser96}
A.~von Mayrhauser and A.~M. Vans, \enquote{\textsl{On the role of hypotheses
  during opportunistic understanding while porting large scale code}}. In 4th
  \textit{Workshop Program Comprehension}, pp.\ 68--77, Mar 1996,
  \doi{10.1109/WPC.1996.501122}.

\bibitem{vonmayrhauser98}
A.~von Mayrhauser and A.~M. Vans, \enquote{\textsl{Program understanding
  behavior during adaptation of large scale software}}. In 6th \textit{Workshop
  Program Comprehension}, pp.\ 164--172, Jun 1998,
  \doi{10.1109/WPC.1998.693345}.

\bibitem{vonmayrhauser97}
A.~von Mayrhauser, A.~M. Vans, and A.~E. Howe, \enquote{\textsl{Program
  understanding behavior during enhancement of large-scale software}}.
  \textit{J.\ Softw.\ Maintenance: Res.\ \& Pract.} \textbf{9(5)}, pp.\
  299--327, Sep/Oct 1997,
  \doi{10.1002/(SICI)1096-908X(199709/10)9:5$<$299::AID-SMR157$>$3.0.CO;2-S}.

\bibitem{weissman74}
L.~Weissman, \enquote{\textsl{Psychological complexity of computer programs: An
  experimental methodology}}. \textit{SIGPLAN Notices} \textbf{9(6)}, pp.\
  25--36, Jun 1974, \doi{10.1145/953233.953237}.

\bibitem{wilson19}
L.~A. Wilson, Y.~Senin, Y.~Wang, and V.~Rajlich, \enquote{\textsl{Empirical
  study of phased model of software change}}, Apr 2019. ArXiv:1904:05842
  [cs.SE].

\bibitem{wohlin:exp}
C.~Wohlin, P.~Runeson, M.~{H\"ost}, M.~C. Ohlsson, B.~Regnell, and
  A.~Wessl\'en, \textit{Experimentation in Software Engineering}.
  Springer-Verlag, 2012.

\end{thebibliography}

\end{document}